\documentclass{osa-article}

\journal{osac}
\usepackage{nicefrac}

\articletype{Research Article}
\usepackage{float}

\begin{document}

\title{Camera detection and modal fingerprinting of photonic crystal nanobeam resonances}

\author{Francis O. Afzal, Joshua M. Petrin, and  Sharon M. Weiss}

\address{\authormark{}Department of Electrical Engineering and Computer Science, Vanderbilt University, Nashville, TN 37235, USA\\}

\email{\authormark{*}francis.afzal@vanderbilt.edu} 



\begin{abstract}
We demonstrate in simulation and experiment that the out-of-plane, far-field scattering profile of resonance modes in photonic crystal nanobeam (PCN) cavities can be used to identify resonance mode order. Through detection of resonantly scattered light with an infrared camera, the overlap between optical resonance modes and the leaky region of k-space can be measured experimentally. Mode order dependent overlap with the leaky region enables usage of resonance scattering as a "fingerprint" by which resonant modes in nanophotonic structures can be identified via detection in the far-field. By selectively observing emission near the PCN cavity region, the resonant scattering profile of the device can be spatially isolated and the signal noise introduced by other elements in the transmission line can be significantly reduced, consequently improving the signal to noise ratio (SNR) of resonance detection. This work demonstrates an increase in SNR of $\sim 19$ dB in out-of-plane scattering measurements over in-plane transmission measurements. The capabilities demonstrated here may be applied to improve characterization across nanophotonic devices with mode-dependent spatial field profiles and enhance the utility of these devices across a variety of applications.
\end{abstract}

\section{Introduction}
Nanophotonic devices such as ring resonators, photonic crystals, plasmonic structures and Fabry-Perot cavities enable numerous optical applications, including signal processing \cite{grapheneringres,RaineriPaper4_NatPhot2017,SubwavelengthPlasmonicLaser_Nature2009}, optical nanomanipulation \cite{plasmonictrapping,EricksonNanomanipulation}, sensing \cite{WhisperingGallerySensors_AIOP2015,SingleNanoparticleDetection,SPRMetalDichalconidesandSi_SciRep2016}, quantum computing \cite{NVCenterDiamonPCNLoncar_NanoLett2013,QubitRingResSi_NatComm2015} and optomechanics \cite{OptomechChaosTransferWhisperingGallery_NatPhot2016,slicedphc1,TrampolineResonatorReview_OpEx2011}, in platforms compatible with dense integration and on-chip implementation. The appeal of these devices largely hinges on their ability to spatially and temporally localize light. As this ability of nanophotonic devices to confine light in time (given by the quality factor, $Q$) and space (given by the mode volume, $V_m$) is dependent on the properties of their available optical modes, it is crucial that optical modes be reliably identified so the appropriate mode(s) may be utilized for a given application.


Currently, experimental identification of mode order in nanophotonic devices has been done by either relating measured modal features (such as resonant peaks in transmission or reflection) to simulated results \cite{DeterministicDesign,FPPlasmonic_JOSAB2012}, or through direct, near-field measurements such as near-field scanning optical microscopy (NSOM) \cite{ShurenArxiv}. While comparison of measured modal features to simulated results is widely used for determining mode order, devices with high spatial or temporal optical confinement are highly sensitive to small changes in device dimensions such that small deviations between simulated and fabricated geometric dimensions can lead to large shifts in resonant features. Furthermore, in some cases, the lowest order modes may not be measurable \cite{SamPaper}. Hence, the comparison between simulation and measurements for mode identification is less reliable for devices with low $V_m$ or high $Q$. NSOM is able to circumvent these issues through direct field measurements of excited optical modes; however, it is a relatively complex technique which is not yet widely available for measurement of guided wave photonic structures.


This work aims to show it is possible to gain information about the mode order of nanophotonic devices by taking advantage of the portion of the mode which couples to free-space propagation and then utilizing the resulting resonant scattering measured in the far-field as a "fingerprint" for identifying mode order. To this end, we analyze the out-of-plane scattering from a photonic crystal nanobeam (PCN) cavity with a standard infrared (IR) camera and demonstrate identification of mode order via modal fingerprinting. We additionally show that it is possible to realize an improved signal-to-noise ratio for PCN resonance features from camera measurements. Utilization of a camera instead of a single photodetector enables spatial isolation of resonance features from noise in the optical signal line. The spatial information provided by camera measurements allows for improved resonance characterization and higher sensitivity measurements of PCN cavities when fabrication imperfections generate significant signal noise. We note that previous work has shown IR cameras may be used for sensing in coupled-resonator optical waveguide (CROW) systems \cite{CROWIRcamerabiosensor_SciRep2014}, characterizing optical propagation in side-coupled integrated spaced sequence of resonators (SCISSOR) and photonic crystal waveguides \cite{SCISSORcharacterizationIRCamera_OpticsLett2010,PhCwaveguideIn-PlaneDisorder_OpEx2009}, and detecting scattering from Fabry-Perot and photonic crystal cavity modes \cite{MidIRPhCSiIRCamera_OpEx2011,JudsonSlot}.


\section{Overview of Technique}
\subsection{k-Space and Modal Fingerprints}
\begin{figure}[b]
\center
\includegraphics[scale = .4,trim={1.8cm 2.3cm 2.0cm 2.5cm},clip,page=1]{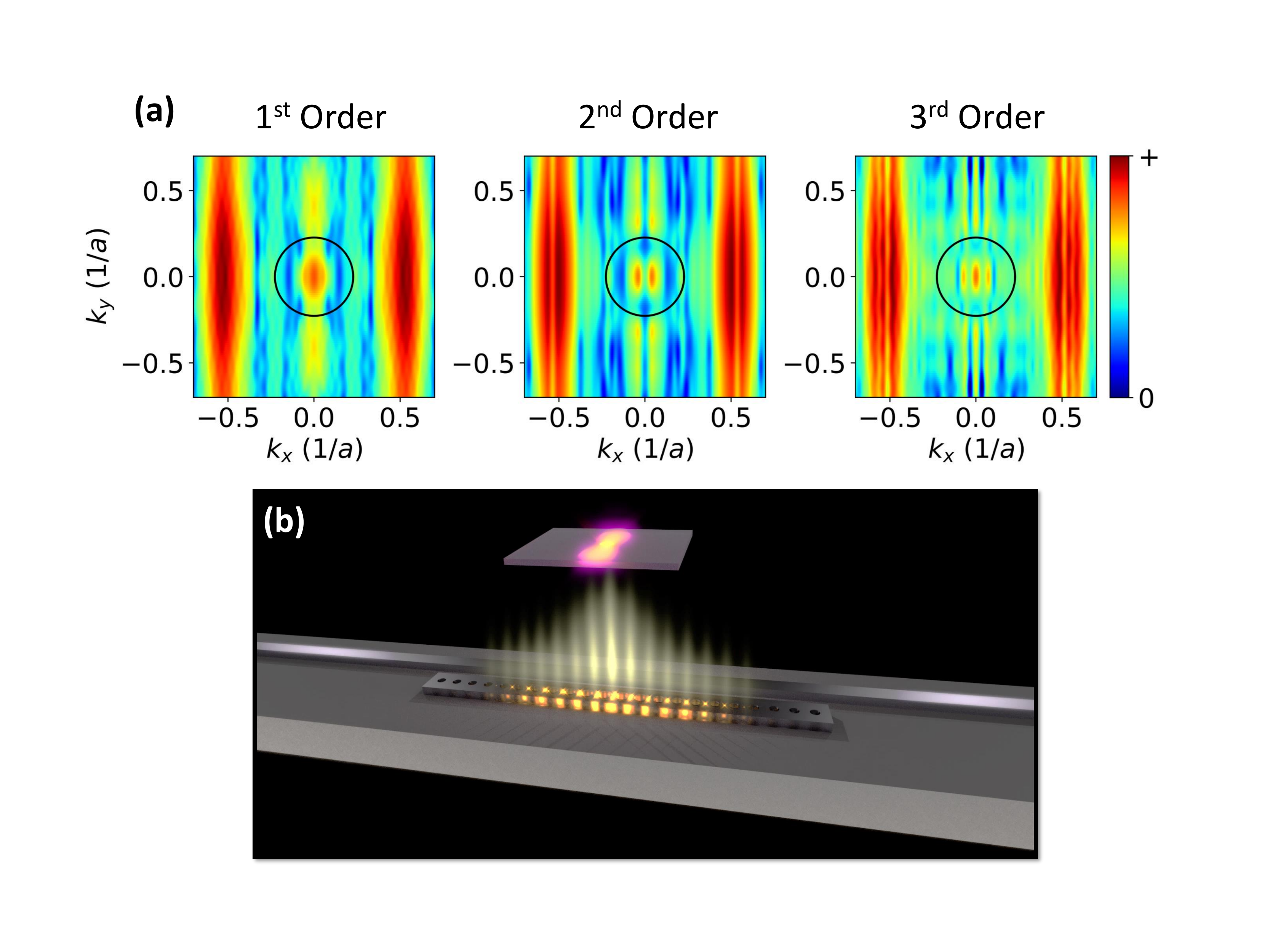}
\caption{(a) k-space profiles for the first three resonance mode orders in a PCN cavity. The leaky region of k-space is enclosed in the black circles. Within the leaky region, we observe the mode shape changes with mode order. (b) Illustration of detection scheme. Resonance mode excitation in PCN cavities leads to out-of-plane scattering according to the mode shape in the leaky region. Excitation of different mode orders produces different scattering patterns in the far-field that can be detected by a camera.} 
\label{Fig:FingerprintandIllustration}
\end{figure}

Optical modes of varying order in nanophotonic devices generally have distinct spatial field profiles and, consequently, have distinct profiles in k-space. The portion of the optical modes in k-space which overlap with k-vectors supported in free-space can be used to directly examine modal properties in the far-field. This leaky region of the mode in k-space, which can be measured in the far-field, may then act as a readily detectable "fingerprint" by which mode order can be identified for a wide range of nanophotonic devices, including PCN cavities. The k-space profiles of the first three resonance mode orders in a PCN cavity are shown in Fig. \ref{Fig:FingerprintandIllustration}(a). The design of this PCN cavity is specified in Sec. \ref{subsec:specifications}.

Figure \ref{Fig:FingerprintandIllustration}(b) illustrates that the only requirements to measure the modal fingerprint of a PCN resonance mode are to simply excite the resonant mode and then measure the scattering in the far-field with a camera. To excite PCN resonances, this work utilizes evanescent coupling (see Sec. \ref{sec:DesignandSim}), but we note it is also possible to fingerprint PCN modes using an in-line coupling configuration under certain circumstances (see Sec. \ref{Appendix:InLineFingerprint} in the Appendix).

\subsection{Considerations for Out-of-Plane Detection}
\label{subsec:outofplanedetection}
For the detection of out-of-plane resonance scattering in photonic devices, there are two primary considerations. (1) devices must have sufficient out-of-plane scattering and (2) non-resonant scattering should be minimized. In general, most fabricated, resonant photonic devices exhibit out-of-plane resonant scattering due to overlap of the resonance mode with the leaky region of k-space. Even high $Q$ cavities that have been designed to minimize overlap with the leaky region in k-space may experimentally demonstrate out-of-plane scattering due to a non-zero overlap with the leaky regions in k-space and fabrication imperfections increasing the overlap of the resonance mode with the leaky region.

Measurement of resonance scattering profiles requires that the resonance scattering intensity is sufficiently large compared to the background signal noise. The intensity of background scattering will largely be dependent on the method of resonance excitation. In-plane, edge-coupling to resonant devices such as Fabry-Perot resonators or photonic crystal cavities can introduce large scattering intensities at the interface between the resonator and the feeding optical mode if the coupling efficiency to the resonator is low. Out-of-plane, free-space coupling to resonators will also introduce a significant amount of non-resonant scattering and reflections at the surface of the device, but this may be filtered out with the proper use of polarizers and resonator orientation \cite{HighQoutofPlanenanobeam,2DfanoHighQ}. In-plane, evanescent coupling to resonant modes with a low-loss waveguide will generally produce low background scattering as energy is coupled between devices via evanescent tails with low field intensity.


\section{Design and Simulations}
\label{sec:DesignandSim}
\subsection{Side-coupled PCN Cavity Specifications}
\label{subsec:specifications}
To demonstrate modal fingerprinting using a camera, we choose to test an evanescently side-coupled PCN cavity because it is relatively straightforward to minimize non-resonant scattering in such a structure \cite{KspaceNanobeamPaper_ArXiv2018}. For the PCN cavity tested in this work, we utilized a cavity taper length of 20 unit cells from the center of the cavity to the edge of the mirror regions and mirror regions of 10 extra unit cells. Devices were designed for fabrication on a 220nm thick device layer on silicon-on-insulator (SOI) wafers with a 3 $\mu\mathrm{m}$ buried oxide layer (BOX) for insulating the cavity mode. A hole period of $350$ nm and PCN width of 700 nm were utilized to move the resonance mode away from the light cones in air and glass. The mirror strength curve utilized for the unit cells in the cavity taper was computed from MIT Photonic Bands (MPB) simulations and is shown in Fig. \ref{Fig:SchematicAndTransmission}(a). We linearly tapered the hole radius from 127nm in the center of the cavity to 108nm at the edges of the cavity. To adequately side-couple into resonance modes, we utilize a coupling gap of 400nm and straight bus waveguide width of 410nm for appropriate overlap of bus waveguide and cavity modes in both physical and k-space \cite{KspaceNanobeamPaper_ArXiv2018}. 

\subsection{Addressing Considerations for Out-of-Plane Detection}
\begin{figure}[t]
\center
\includegraphics[scale = .36,trim={1cm 3.6cm 1cm 2cm},clip,page=2]{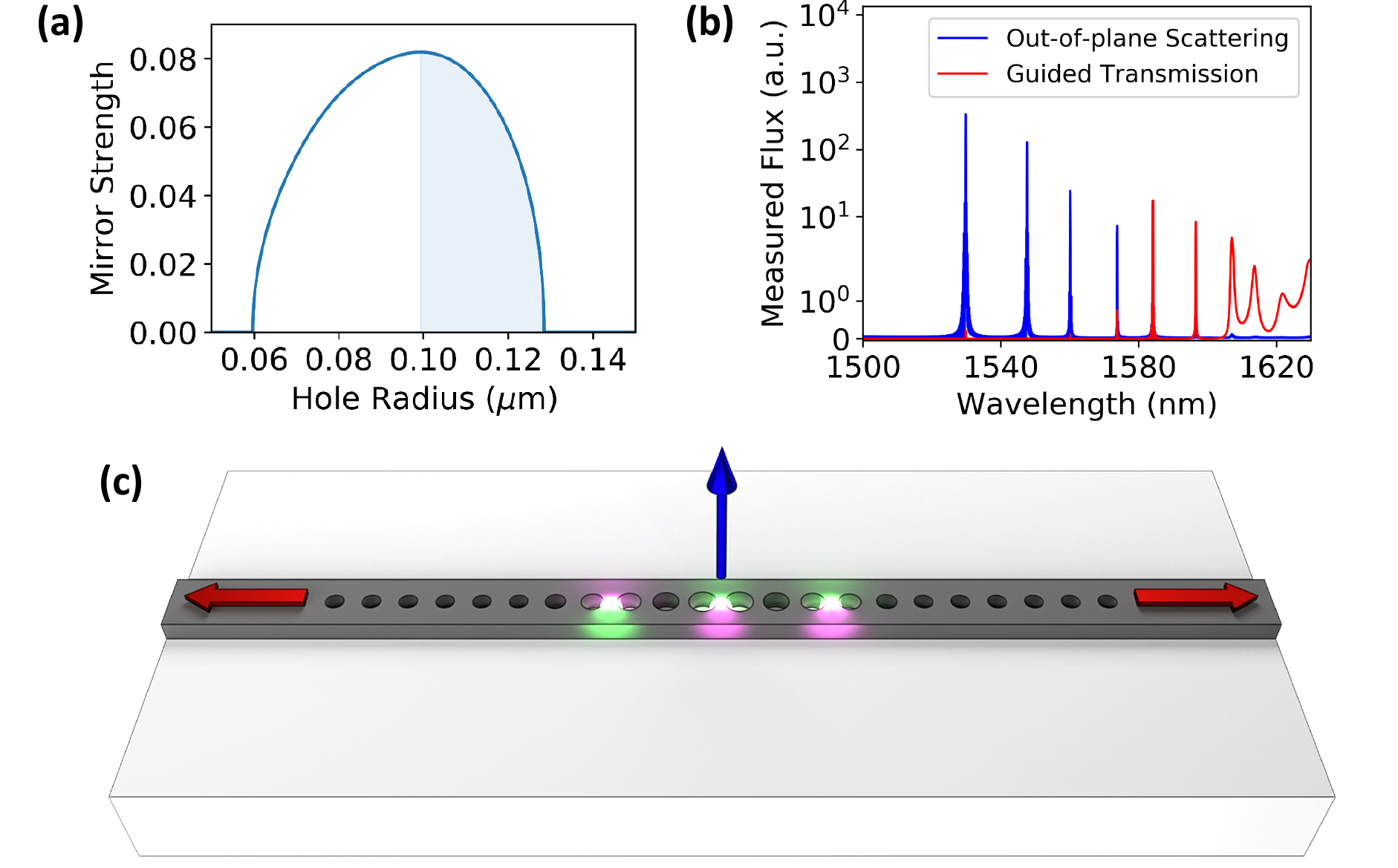}
\caption{(a) Mirror strength curve used in cavity design. The shaded region under the curve corresponds to hole radii used in creating PCN cavities with dielectric mode cavity resonances. (b) Measured flux from out-of-plane scattering (blue) and in-plane guided transmission (red) from PCN resonance excitation with 3 dipole sources. Lower order, more well confined modes are observed to scatter more selectively out-of-plane rather than coupling into guided modes. (c) Illustration of dipole configuration used to excite resonance modes. Actual dipole positions used were at the center of the cavity and $\pm 2.1 \mu \mathrm{m}$ from the center of the cavity. The two outer sources had a $\pi$ phase offset to excite even order modes.} 
\label{Fig:SchematicAndTransmission}
\end{figure}

To (1), verify our designed PCN geometry had adequate out-of-plane scattering on resonance, we simulated resonance mode excitation (specified in Sec. \ref{subsec:simspecs}) and monitored the optical flux both out of plane and at the edge of the cavity. Figure \ref{Fig:SchematicAndTransmission}(b) shows significant out-of-plane scattering on resonance for our PCN design when excited.

To (2), reduce non-resonant scattering, we chose to utilize a side-coupled configuration with a straight bus waveguide to evanescently excite PCN resonance modes. We present guidelines on how to side-couple to PCN cavities with a straight bus waveguide in previous work \cite{KspaceNanobeamPaper_ArXiv2018}. Implementing a straight side-coupled bus waveguide instead of a curved side-coupled bus waveguide reduces scattering due to bending losses in a curved bus waveguide. We note it is possible to observe mode-dependent resonance scattering with an in-line configuration which couples light in directly from the edge of the PCN cavity. Scattering noise from coupling through the edges of the cavity can be reduced by increasing the coupling efficiency to resonance modes, but this generally reduces the loaded $Q$ of resonance modes (see Sec. \ref{Appendix:InLineFingerprint} in the Appendix).

\subsection{Simulation Specifications and Modal Profiles}
\label{subsec:simspecs}
Far-field scattering profiles of resonance modes were computed via 3D calculations in Lumerical FDTD Solutions. These simulations were run for 6000fs simulation time and with an auto non-uniform meshing factor of 5. To computationally excite PCN resonance modes, we utilized 3 dipole sources placed at different points in the PCN cavity, as illustrated in Fig. \ref{Fig:SchematicAndTransmission}(c). Two dipole sources were placed at $\pm 2.1 \mu \mathrm{m}$ from the cavity center with separate phases of $0$ and $\pi$. Another dipole source was placed at the center of the cavity with a phase of $0$. The phase offset between the non-central sources was utilized to excite even order modes in the PCN while the central dipole source was used to excite odd order modes. The placements of the sources were chosen to tune the spatial field overlap between the dipole sources and the modes into which they constructively coupled.

\begin{figure}[t]
\center
\includegraphics[scale = .36,trim={0.1cm 5.5cm 0.1cm 2.5cm},clip,page=3]{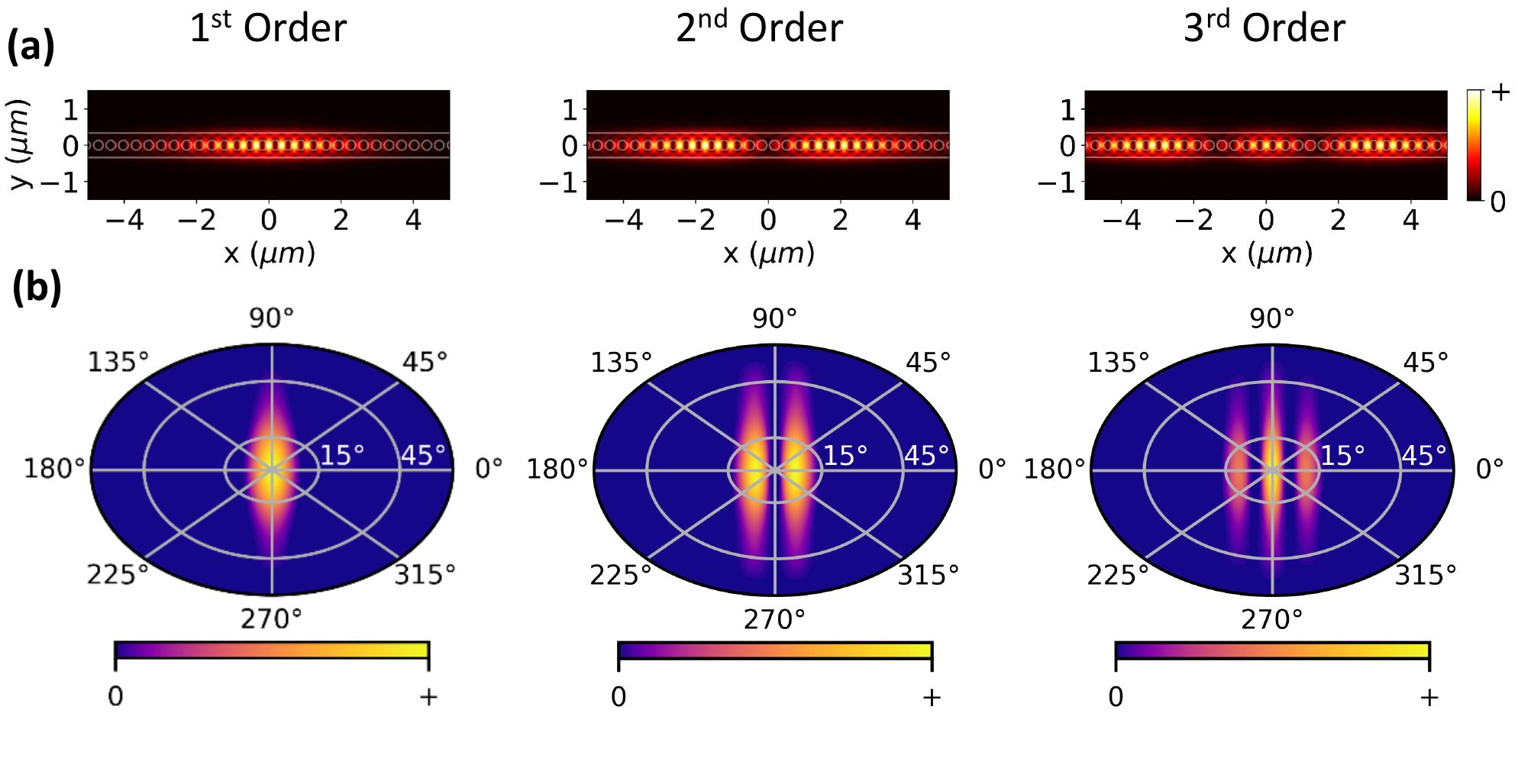}
\caption{(a) Spatial field profiles for the first three resonance modes in the designed PCN cavity. (b) Calculated scattering in the far field for the first three resonance modes. It is clear that each near field profile in (a) is uniquely connected to a far-field pattern in (b).} 
\label{Fig:NearandFarField}
\end{figure}

The spatial field profiles of the first three PCN cavity modes are illustrated in Fig. \ref{Fig:NearandFarField}(a). To calculate the far-field of the resonance modes, a field monitor was utilized to capture the electric and magnetic field components of the resonance modes $\sim 170 \mathrm{nm}$ above the cavity. The far-field scattering profile was then extrapolated using the near field components and propagating them into the far-field using standard FDTD methods \cite{NanophotonicPhasedArray_Nature2013, FDTDBook}. The calculated far-field scattering profiles are shown in Fig. \ref{Fig:NearandFarField}(b). The far field patterns for each resonance mode simulated are uniquely dependent on the mode order and are similar in shape to the k-space profiles of each mode in the leaky region shown in Fig. \ref{Fig:FingerprintandIllustration}(a). The dependence of scattering on mode order enables the use of resonance scattering patterns as "fingerprints" for identifying resonance modes. Scattering patterns from higher order modes are discussed in Sec. \ref{Appendix:HighOrderFingerprint} in the Appendix.

\section{Experiments}
\subsection{Fabrication Method and Characterization Setup}
To experimentally demonstrate modal fingerprinting and compare scattering and transmission line measurements, we fabricated side-coupled PCN cavities using electron-beam lithography (EBL) and reactive ion etching (RIE). Devices were fabricated on chips from a silicon-on-insulator (SOI) wafer with a $220 \mathrm{nm}$ thick silicon device layer and a $3 \mathrm{\mu m}$ thick buried oxide layer (SOITEC). To pattern devices on the chips, we spin-coated 300nm of ZEP520A photoresist at 6000rpm for 45s and exposed the resist using a JEOL93000FS EBL tool. The photoresist was patterned with an electron beam at 100kV and 400 $\mathrm{\mu C/cm^{2}}$ areal dosage. Patterns were developed after exposure with gentle agitation in xylenes for 30s followed by a rinse in isopropyl alcohol. The photoresist pattern was then transferred to the SOI device layer via RIE using an Oxford PlasmaLab100. RIE was carried out with C$_{4}$F$_{8}$/SF$_{6}$/Ar gases. Samples were cleaved after fabrication to expose the edges of the feeding waveguides to enable characterization in an end-fire transmission setup. An SEM image of a fabricated, side-coupled PCN cavity is shown in Fig. \ref{Fig:ExperimentalData}(a).

The transmission spectra of the side-coupled PCN cavities were measured by coupling near-infrared light from a tunable laser (1500 to 1630nm, Santec TSL-510) into and out from the bus waveguides using polarization-maintaining lensed fibers (OZ Optics Ltd.), and detecting the output light intensity with a fiber-coupled avalanche photodiode photoreceiver (Newport 2936-C). TE polarized light was utilized in the experiments.

The scattered optical profile was measured in an InGaAs camera (New Imaging Technologies WiDy SWIR 640U-S) through a 20x, large working distance (20 mm) objective (Mitutoyo Plan Apo NIR) with a numerical aperture (NA) of 0.40. We note that an essential element in detecting resonant cavity scattering requires matching the cavity scattering angles with the measurable angle range associated with the NA of the implemented objective. As much of the simulated scattering depicted in Fig. \ref{Fig:NearandFarField}(b) occurs within $15^{\circ}$ of normal incidence, we found it sufficient to use NA = 0.40 in this work, which corresponds to an acceptance angle of $\sim 24^{\circ}$ for the objective. 

Utilizing the end-fire setup to couple light into a bus waveguide, we evanescently excite the cavity resonance mode and measure the scattered light with an IR camera. The captured video from the camera was processed with ImageJ to spatially isolate the scattering around the device for analyzing average scattering intensity as a function of wavelength. We additionally measure light transmitted through the bus waveguide for verification of resonance wavelength position and performance comparison. No extra polarizers or filters are used in our camera measurement. Objectives with larger NA could be used to resolve higher order resonance scattering from higher scattering angles.


\subsection{Experimental Data}
\begin{figure}[t]
\center
\includegraphics[scale = .36,trim={0.1cm 0cm 0.2cm 0cm},clip,page=4]{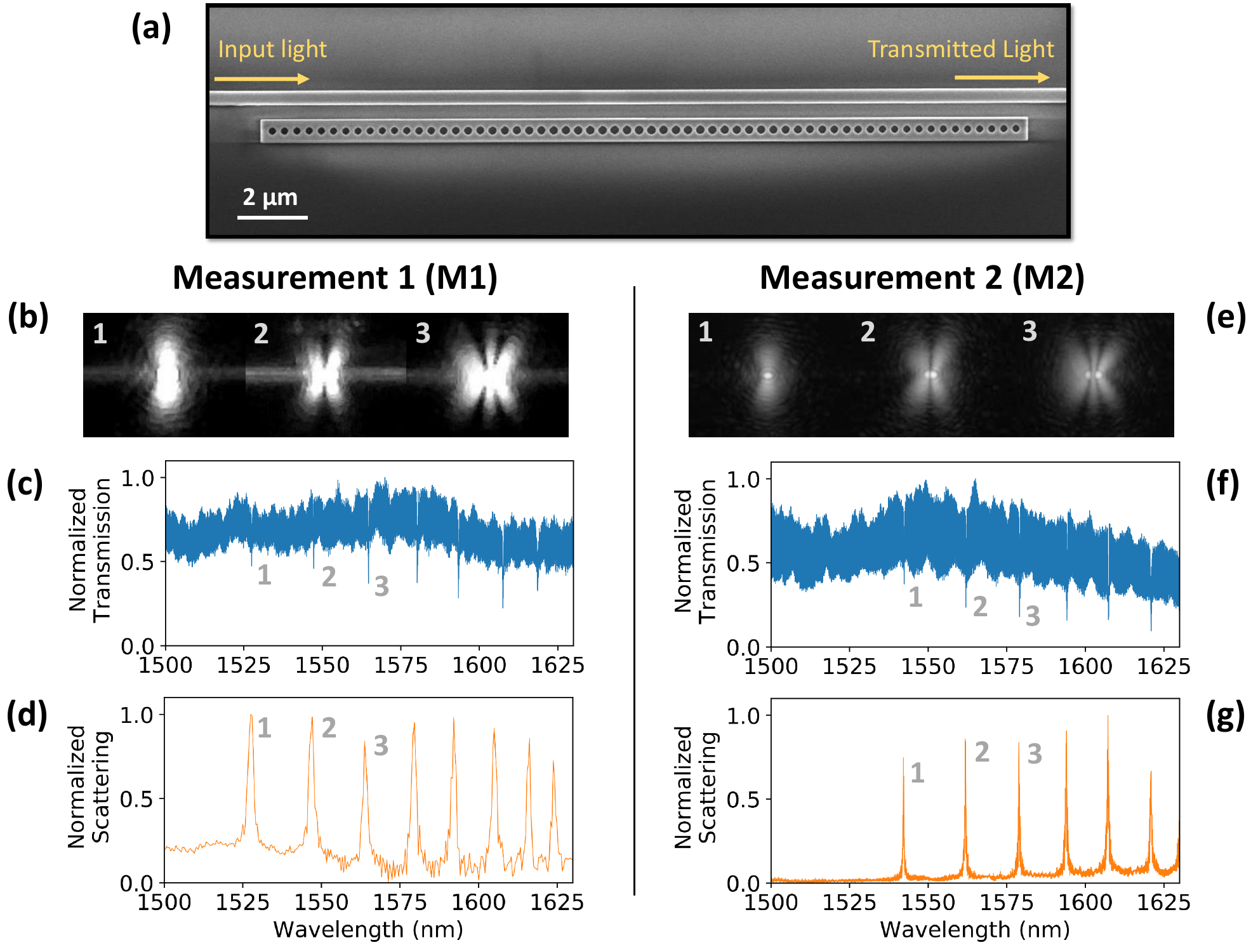}
\caption{(a) SEM image of fabricated, side-coupled PCN cavity. The optical input from the end-fire setup feeds the bus waveguide. The transmitted light through the bus waveguide is sent to a detector to characterize resonance peaks. The out-of-plane resonantly scattered light from the cavity is captured with an IR camera (see Visualization 1 for video data used in M2). (b) Scattered optical profiles recorded by an IR camera for resonances labeled (1), (2) and (3) for M1. The resonance spectrum was measured via (c) transmission through bus waveguide and (d) average scattering detected by IR camera for M1. (e) Scattered optical profiles for resonances labeled (1), (2) and (3) in M2. The resonance spectrum was measured via (f) transmission through bus waveguide and (g) average scattering detected by IR camera for M2. A higher data point density and modified digital detection range were used for M2.} 
\label{Fig:ExperimentalData}
\end{figure}

Experimentally measured resonant scattering profiles, transmission spectra, and scattered optical intensities for the fabricated PCN are shown in Fig. \ref{Fig:ExperimentalData}.  For clarity, it is important to explain that the data in Fig. \ref{Fig:ExperimentalData}(b-d) and Fig. \ref{Fig:ExperimentalData}(e-g) are from the same physical device, but have been taken with different data collection parameters and with different levels of surface adsorbed water, which alter the measurement quality and position of resonance peaks, respectively. Accordingly, data from Fig. \ref{Fig:ExperimentalData}(b-d) will be referred to as measurement 1 (M1) and Fig. \ref{Fig:ExperimentalData}(e-g) will be referred to as measurement 2 (M2) in this paper. With this data, we demonstrate (I) modal fingerprinting of resonances, (II) scattered optical profile tolerance to resonance wavelength shifts and (III) improvement in SNR from spatial isolation of resonance structure. We also show how measurement considerations such as (i) data point density and (ii) digital detection range can be used to improve measurement quality. 

To (I), demonstrate modal fingerprinting of resonances, we measured the transmission spectrum of side-coupled PCN devices while simultaneously recording video data of the scattered light. As the wavelength of the input laser light was swept, dips appeared in the transmission spectrum on resonance (Fig. \ref{Fig:ExperimentalData}(c,f)). At these resonance wavelengths, we additionally observed increased scattering in the video data from the camera (see Visualization 1 in Fig. \ref{Fig:ExperimentalData}). The peaks in scattering intensity shown in Fig. \ref{Fig:ExperimentalData}(d,g) correspond directly with dips in the transmission spectra shown in Fig. \ref{Fig:ExperimentalData}(c,f). As seen in Fig. \ref{Fig:ExperimentalData}(b,e), the first three resonance modes imaged by the camera yield unique scattering profiles, demonstrating that the scattered optical profile may be utilized to identify resonance mode order in these devices. Experimentally measured scattering profiles of higher order modes are discussed in Sec. \ref{Appendix:HighOrderFingerprint} in the Appendix.

In order to (II), show the scattered optical profile is tolerant to resonance peak shifts, we measured side-coupled PCN devices before (M1) and after (M2) prolonged exposure to ambient humidity in an unregulated environment. We utilized the exposure to ambient humidity to induce adsorption of water on the surface of the devices and increase the refractive index of the cavity, red-shifting the resonance wavelengths, as shown by comparing Fig. \ref{Fig:ExperimentalData}(c,d) and Fig. \ref{Fig:ExperimentalData}(f,g). Because this prolonged humidity exposure was unregulated, we simply use it to qualitatively show the resonance scattering pattern is preserved upon resonance peak shifting, as the resonance profile in real and k-space is not significantly perturbed. While the pixel ranges and saturation settings are different between M1 and M2, we see that the shape of resonance scattering in Fig. \ref{Fig:ExperimentalData}(b,e) is clearly preserved between the two measurements. We note that the $\sim 15$nm resonance red-shift between M1 and M2 from extended exposure to ambient atmospheric conditions occurs within a reasonable range when compared to previous results on photonic crystal cavities used as humidity sensors \cite{Humidity1_OptLett2016,Humidity2_APL2013}.

We additionally show that (III), improvement in SNR from spatial isolation of the PCN resonance structure can result from implementing camera detection of resonant scattering. Noise in transmission line measurements from Fig. \ref{Fig:ExperimentalData}(c,f) results from Fabry-Perot interference caused by reflections from the waveguide facets at the edges of the end-fire coupling setup, waveguide edge roughness, and stitching errors which occur during EBL. In general, it is possible to reduce this noise in the transmission line by various means, but in practice this can be difficult and requires further development of fabrication processes. By utilizing a camera to detect local resonance scattering, we can spatially isolate the PCN cavity from imperfections and reflections in the bus waveguides. This results in an improved SNR relative to the transmission line measurements. In the data from M2 where we have improved our scattering measurement, Fig. \ref{Fig:ExperimentalData}(f,g), we calculate the SNR of the fundamental resonance mode to be 2.10 dB from transmission line measurements and 21.4 dB from measurement of scattered light. These SNR values were calculated by considering the resonance peak height with respect to the standard deviation of the local signal baseline. A $Q$ of $\sim 6 \times 10^3$ was measured for the fundamental resonance by both techniques in M2. A comparison of the two SNR values from the two methods shows a 19.3 dB increase in the SNR by utilizing the scattered light to measure resonance peaks. We attribute this improvement to camera measurements enabling the spatial isolation of cavity scattering from guided and un-guided scattering in the feeding waveguides.

To improve the measurement quality of scattered optical profiles and resonance peak widths with the camera, we altered (i) data point density and (ii) digital detection range between M1 and M2. To (i) alter the data point density in the spectrum measured by the camera, the wavelength scan rate of the tunable laser source and the captured frames per second (fps) of the camera were adjusted. In M1 (Fig. \ref{Fig:ExperimentalData}(d)), a laser scan rate of 10nm/s and a 24fps constant camera capture rate were utilized. In M2 (Fig. \ref{Fig:ExperimentalData}(g)), a laser scan rate of 1nm/s was utilized and each video frame capture was triggered by the laser at 0.04nm intervals, giving an average frame rate of 25fps. The reduced laser scan rate and higher frame rate of the camera enabled much higher point density in the measured spectrum. The difference in frame-rate and scan speed results in M1 having a spectral resolution of 0.42nm while M2 has a spectral resolution of 0.04nm. Further slowing of the scan rate can be achieved, but this can lead to increasingly large data files as each data point corresponds to a camera image. The uncompressed video file from the camera utilized for extracting the spectrum in M2 had a file size of $\sim$ 1.5 GB. More advanced spatial selection during data capture or file compression could reduce this data requirement.

The (ii) digital detection range of the camera measurements were also adjusted between M1 and M2 to improve measurement quality. The automatic, real-time adjusted image settings used in M1 show clipping behaviour in the images of resonance scattering profiles (Fig. \ref{Fig:ExperimentalData}(b)). This reduces the contrast in the scattered profile and potentially hinders resolution of resonance peak shape in the measured spectra shown in Fig. \ref{Fig:ExperimentalData}(d). To reduce clipping behavior, we manually adjusted the data capture range of the camera to make the maximum pixel capture value high enough to reduce clipping and the minimum pixel capture value low enough to not filter out resonant scattering. This results in data in M2 (Fig. \ref{Fig:ExperimentalData}(e)) showing more contrast and clarity in the resonant scattering image compared to M1.

\section{Conclusions}
The performance of nanophotonic devices hinges on the properties of their supported optical modes. As the different optical modes of a nanophotonic device have different properties and performance metrics, it is crucial to be able to identify different mode orders so the appropriate mode(s) may used in applications. This is particularly true for photonic crystals, as their resonance $Q$, $V_{m}$ and spatial field profiles achieved are highly dependent on mode order. We demonstrate it is possible to utilize mode dependent overlap with the leaky region as a "fingerprint" to identify mode order in nanophotonic devices by analyzing the far-field scattering from PCN cavity resonances. In this work, an IR camera and a side-coupled PCN are implemented to show (I) modal fingerprinting of resonances, (II) scattered optical profile tolerance to resonance peak shifts and (III) improvement in SNR from spatial isolation of resonance structure. It is also shown that measurement considerations such as (i) data point density and (ii) digital detection range can be used to improve measurement quality. Because PCN modes can be accessed by both in-line or side-coupled excitation, modal fingerprinting can identify modes in in-line devices either directly from device scattering or indirectly via scattering of a comparable test device which is either side-coupled or has high coupling efficiency for in-line excitation (see Sec. \ref{Appendix:InLineFingerprint} in the Appendix). By demonstrating modal fingerprinting via resonance scattering, we lower the technical threshold for more comprehensive characterization of modes in nanophotonic devices and leveraging the desired modal properties for targeted optical applications. 

\appendix
\section{Appendix}
\label{appendix}

\subsection{Modal Fingerprinting in In-Line PCN Cavities}
\label{Appendix:InLineFingerprint}
\begin{figure}[b]
\center
\includegraphics[scale = .36,trim={0cm 0cm 4cm 0cm},clip,page=5]{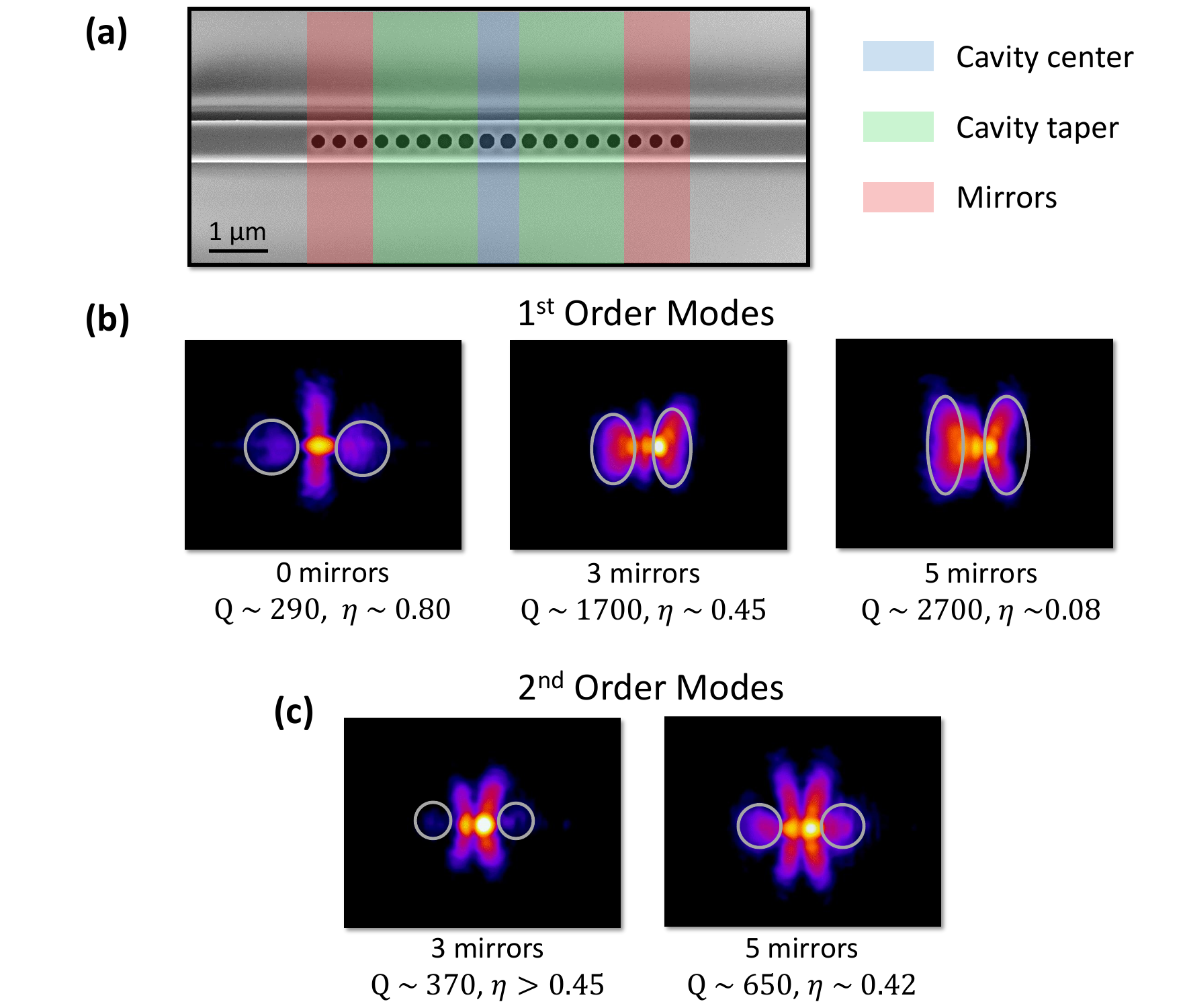}
\caption{SEM of in-line PCN and experimental images of resonant scattering from in-line PCN cavities with no side-coupler present. Excess scattering from the edges of the cavity are circled in grey. False color is used to enhance contrast in the scattering images. Loaded $Q$ and approximate $\eta$ values are reported for resonances in (b) and (c). (a) SEM of a measured PCN cavity in an in-line coupling configuration. The blue, green and red highlighted regions correspond to the cavity center, cavity taper, and extra mirrors added, respectively. (b) Images of scattering from first order modes in separate devices with different numbers of extra mirrors added after the cavity taper. (c) Images of scattering from second order modes in separate devices with different numbers of extra mirrors added after the cavity taper. } 
\label{Fig:InLine}
\end{figure}

Modal fingerprinting can be utilized for PCN cavities in an in-line configuration, although identification of resonance modes is more challenging in cases when there is significant scattering from the edges of the cavity. Figure \ref{Fig:InLine} shows experimental resonant scattering profiles for PCN cavities in an in-line coupling configuration where light is coupled into resonance modes through the edge of the cavity. By varying the number of extra mirrors after the cavity taper, the confinement of resonance modes is altered and consequently changes the coupling efficiency for each mode. The coupling efficiency, $\eta$, for a PCN in an in-line coupling configuration is given by

\begin{equation}
\eta = \mathrm{T_{res}}= \frac{\left(\nicefrac{Q_{0}}{Q_{c}}\right)^2}{\left(1+\nicefrac{Q_{0}}{Q_{c}}\right)^2} \:,
\end{equation}
where $\mathrm{T_{res}}$ is the transmittance on resonance, $Q_{0}$ is the intrinsic quality factor of the resonator and $Q_{c}$ is the coupling quality factor of the resonator. Because $\eta$ is related to the ratio $\nicefrac{Q_{0}}{Q_{c}}$, PCN resonances with high $Q_{c}$ relative to $Q_{0}$ demonstrate low $\eta$. Resonance modes with high $Q_{loaded}$ often have high $Q_{c}$ and low $\eta$, making them difficult to identify by modal fingerprinting in an in-line configuration.

An SEM image of one of the in-line coupled PCN cavities that was measured using the far-field scattering technique is shown in Fig. \ref{Fig:InLine}(a). All of the in-line PCN cavities that were measured were designed with hole tapering from a hole radius of 127 nm at the cavity center to 112 nm at the edges of the cavities with a taper length of 5 unit cells. All other dimensions of the nanobeam are identical to the side-coupled devices discussed earlier. The number of mirror unit cells at each end of the PCN was varied to modify the coupling efficiency. The larger radius at the cavity edges and smaller cavity taper were utilized to lower $Q_{c}$ and non-resonant scattering from the cavities. To estimate $\eta$, we compared the transmitted power of PCN cavities on resonance to the transmission through a reference waveguide. Fig. \ref{Fig:InLine}(b,c) shows that the edge scattering (the circled grey regions) tends to increase with reduced $\eta$. 

To circumvent the difficulty of modal fingerprinting in-line, high $Q$ (and low $\eta$) cavities, it is possible to identify resonances without direct modal fingerprinting of the cavity. The peak positions and modal fingerprints of a side-coupled cavity could be used to identify modes of an in-line cavity with the same geometry, as side-coupling does not significantly perturb the peak positions or modal fingerprint in the weak coupling regime. An in-line device with fewer mirrors or taper segments for higher $\eta$ could also be used to identify the modes of the higher $Q$ device so long as the device changes do not significantly alter the general resonance shape and peak positions between the devices.

\subsection{Fingerprinting Higher Order PCN Modes}
\label{Appendix:HighOrderFingerprint}
\begin{figure}[H]
\center
\includegraphics[scale = .36,trim={2cm 3.5cm 2cm 0cm},clip,page=6]{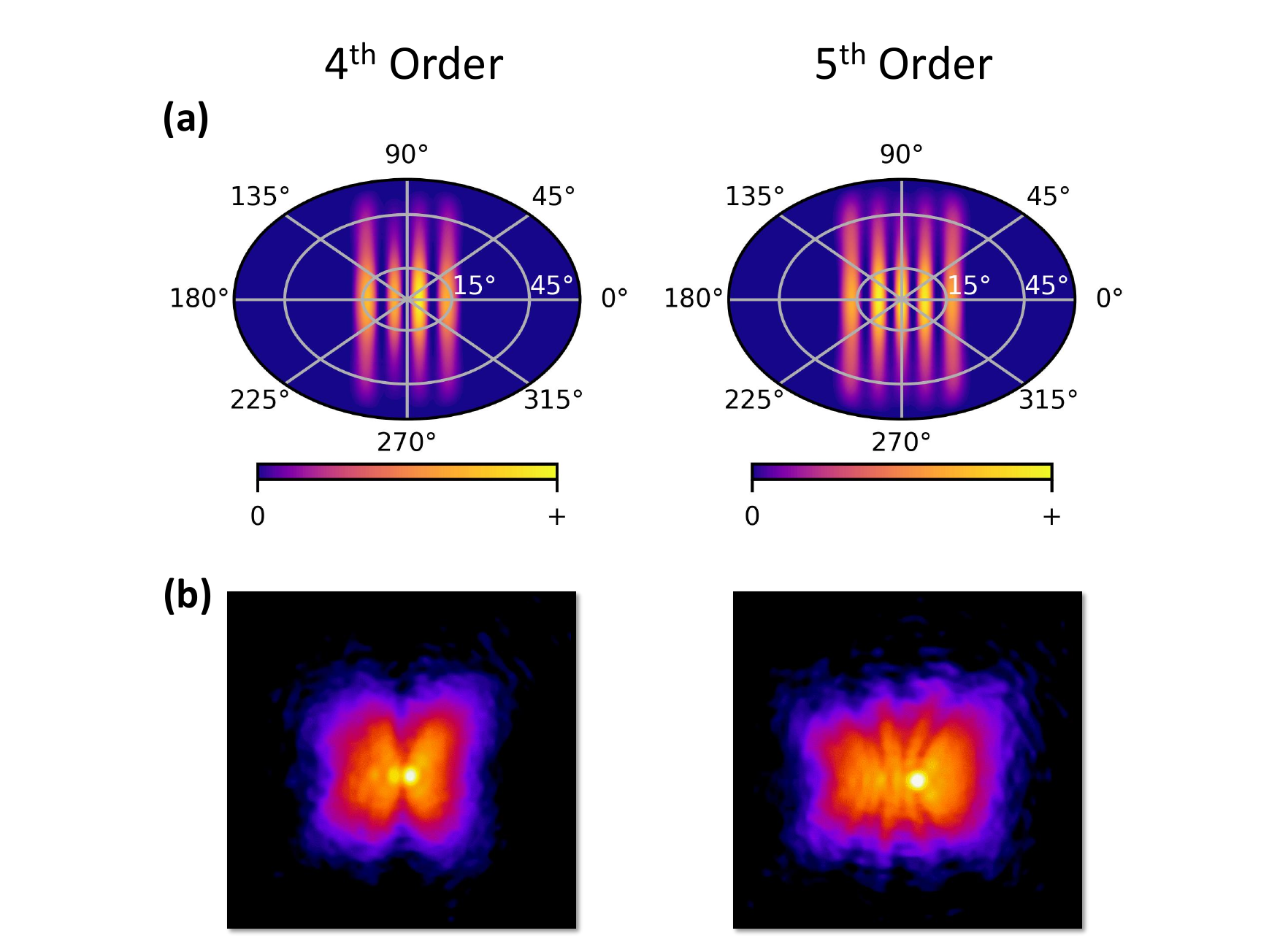}
\caption{Side-by-side comparison of (a) simulated scattering profiles and (b) experimentally measured images of resonance scattering for the fourth and fifth order resonances of the fabricated side-coupled PCN cavity. False color is used in the measured scattering data to enhance contrast in the images.} 
\label{Fig:HighOrderModes}
\end{figure}

The simulated scattering for fourth and fifth order resonance modes in the side-coupled PCN cavity (see SEM image in Fig. \ref{Fig:ExperimentalData}(a)) are shown in Fig. \ref{Fig:HighOrderModes}(a). The tight angular packing and wider angle range of scattered light in these higher order modes makes them more difficult to resolve in the camera images shown in Fig. \ref{Fig:HighOrderModes}(b). Increasing camera resolution ($640 \times 512$ pixels in the camera used here) or utilization of a higher power objective (20x used here) could further improve the ability to resolve higher order modal fingerprints.

\section*{Funding}
National Science Foundation (ECCS1407777 and ECCS1809937); National Science Foundation Graduate Research Fellowship (F. O. Afzal).

\section*{Acknowledgments}
Fabrication of the photonic crystal nanobeam structures was conducted at the Center for Nanophase Materials Sciences, which is a DOE Office of Science User Facility. Scanning electron microscopy imaging was carried out in the Vanderbilt Institute of Nanoscale Science and Engineering. The authors thank Dr. Shuren Hu for helpful discussions. The authors also gratefully acknowledge Dayrl Briggs, Kevin Lester, Dr. Ivan Kravchenko, Dr. Scott Retterer and Dr. Kevin Miller for helpful discussions and assistance in fabrication. 
\section*{Disclosures}
The authors declare that there are no conflicts of interest related to this article.

\end{document}